\documentstyle[pra,aps,twocolumn]{revtex}
\input epsf
\newcommand{\infig}[2]{\begin{center}\mbox{ \epsfxsize #1
                       \epsfbox{#2}}\end{center}}

\newcommand{\be}{\begin{equation}}
\newcommand{\ee}{\end{equation}}
\newcommand{\bea}{\begin{eqnarray}}
\newcommand{\eea}{\end{eqnarray}}
\begin{document}

\draft

\title{Spatial diffusion in a periodic optical lattice: revisiting the Sisyphus effect}

\author{Laurent Sanchez-Palencia${}^1$\footnote{e-mail: lsanchez@lkb.ens.fr}, Peter Horak${}^2$\footnote{present adress:  Optoelectronics Research Centre,
University of Southampton, UK.} and Gilbert
Grynberg${}^1$}

\address{${}^1$ Laboratoire Kastler-Brossel, D\'epartement de Physique de
l'Ecole Normale Sup\'erieure, 24, rue Lhomond, F-75231 Paris cedex
05, France}

\address{${}^2$ Institut f\"ur Theoretische Physik, Universit\"at Innsbruck,
Technikerstra{\ss}e 25, A-6020 Innsbruck, Austria}

\date{\today{}}

\maketitle

\begin{abstract}
We numerically study the spatial diffusion of an atomic cloud experiencing
Sisyphus cooling in a three-dimensional lin$\bot$lin optical lattice in a broad
range of lattice parameters. In particular, we investigate the dependence on
the size of the lattice sites which changes with the angle between the laser
beams. We show that the steady-state temperature is largely independent of the
lattice angle, but that the spatial diffusion changes significantly. It is
shown that the numerical results fulfil the Einstein relations of Brownian
motion in the jumping regime as well as in the oscillating regime. We finally
derive an effective Brownian motion model from first principles which gives
good agreement with the simulations.
\end{abstract}

\pacs{PACS. 32.80.Pj Optical cooling of atoms, trapping - 42.50.Vk
Mechanical effects of light on atoms, molecules, electrons and ions}

\section{Introduction}

Laser cooling and trapping was one of the major advances of the
last part of the 20th century. In 1997, the Nobel prize in physics
was awarded to Steven Chu, Claude Cohen-Tannoudji and William D.\
Phillips for their works in this domain \cite{Nobel97}, in
particular, for their discovery of the Sisyphus cooling effect
\cite{Sisyphus89} which permits to achieve sub-Doppler
temperatures and is widely used in various different laser cooling
schemes \cite{Metcalf99}. The Sisyphus effect in optical lattices
\cite{Jessen96} was studied in a large variety of systems and of
field configurations. A large number of results were obtained on
temperature, localization and spatial order \cite{optlatt2001} and
an excellent agreement between the experimental observations and
the theoretical predictions was found. Much less work has been
done to study the spatial diffusion in optical lattices, but also
for this problem a reasonable agreement was found between the
models \cite{oscillth95,anomalous96,localization96} and the
experiments \cite{oscillth95,Jurkzak96,ions97,Truscott98}.
However, no detailed study of the dependence of spatial diffusion
on, for example, the different directions of an anisotropic
lattice or on the size and shape of lattice sites has been
performed so far.  Very recently, optical lattices and atomic
transport therein has attracted new attention with the study of
quantum chaos \cite{chaos} and the achievement of Bose-Einstein
condensation by purely optical means \cite{Chapman}.

Spatial diffusion in one-dimensional (1D) Sisyphus cooling schemes
is fairly well understood. In the so-called {\it jumping regime},
where an atom undergoes several optical pumping cycles while
moving over one optical wavelength, the atomic motion can be
understood by a simple model of Brownian motion
\cite{Sisyphus89,Cohen90}. In this regime, spatially averaged
friction coefficients $\alpha_0$ and momentum diffusion
coefficients $D_p$ have been derived and the validity of the
Einstein relations \be k_B T = \frac{D_p}{\alpha_0}
\label{TDpalpha} \ee and \be D_s = \frac{k_B T}{\alpha_0},
\label{DsTalpha} \ee where $T$ is the steady-state temperature,
$k_B$ the Boltzmann constant, and $D_s$ the spatial diffusion
coefficient, has been shown. On the contrary, in the so-called
{\it oscillating regime}, where an atom travels over several
optical wavelengths before being optically pumped into another
internal state, the friction force has been shown to be
velocity-dependent \cite{Sisyphus89,Cohen90}, \be F(v) =
\frac{-\alpha_0 v}{1+(v/v_c)^2} \ee where $v_c$ denotes the
capture velocity of Sisyphus cooling. In this situation, an
analytical derivation of the spatial diffusion coefficient has
still be found \cite{oscillth95}, but an interpretation in terms
of a simple Brownian motion no longer works. In particular, it is
found in Ref.~\cite{oscillth95} that the behavior of the spatial
diffusion coefficient as a function of the atom-light interaction
parameters (laser light intensity and detuning) is dramatically
different in the oscillating regime compared to the jumping
regime.

In higher dimensional setups, such a difference of the spatial
diffusion behaviors in the jumping and the oscillating regimes is
expected, too. Moreover, the important difference of the mean free
path of a diffusing atom in these regimes may induce different
behaviors of the spatial diffusion coefficients as a function of
the lattice periods. These are the main issues considered in the
present paper.

In this work we perform a detailed study of spatial diffusion in
the so-called 3D-lin$\bot$lin lattice \cite{3dlinlin}. Using
semiclassical Monte-Carlo simulations we find that equalities of
the form of Eqs.~(\ref{TDpalpha}) and (\ref{DsTalpha}) still hold
in the oscillating regime. This suggests that an interpretation by
a Brownian motion should still be possible. We derive such a model
from basic principles assuming a thermal spatial distribution and
taking into account some specific properties of our optical
lattice and find a good quantitative agreement with the numerical
results. In particular, we calculate an effective friction
coefficient $\alpha$ in a range of parameters containing both the
jumping and the oscillating regimes. We find that in the
oscillating regime, $\alpha$ increases with the lattice beam
intensity and decreases when $|\Delta|$ increases, in strong
opposition to the friction coefficient $\alpha_0$ calculated in
the jumping regime.

Our work is organized as follows. In Sec.~\ref{sisyphus} we
describe the specific laser and atom configuration for the 3D
optical lattice that we consider here and discuss several
important features of the optical potential surfaces. In
Sec.~\ref{physics}, we present a physical picture of spatial
diffusion in periodic optical lattices and we particularly
forecast a dramatic change of the behavior of the spatial
diffusion coefficients not only versus the atom-light interaction
parameters but also versus the geometrical parameters (spatial
lattice periods) when going from the jumping to the oscillating
regime. In Sec.~\ref{theory} we derive an effective Brownian
motion model which we compare in the following sections with the
numerical results on the steady-state temperature
(Sec.~\ref{temperature}) and on the spatial diffusion
(Sec.~\ref{diffusion}). Numerical results on the friction
coefficient are then discussed in Sec.~\ref{friction} and the
validity of the Einstein relations is shown. Finally, we summarize
our results in Sec.~\ref{conclusions}.

%%%%%%%%%%%%%%%%%%%%%%%%%%%%%%%%%%%%%%%%%%%%%%%%%%%%%%%%%%%%%%%%%

\section{Sisyphus cooling in the 3D-lin$\bot$lin configuration}
\label{sisyphus}

%\subsection{The optical lattice configuration}

The Sisyphus effect cools a cloud of multi-level atoms when a
laser field induces spatially modulated optical potentials and
pumping rates in such a way that a moving atom on average climbs
up potential hills before it is optically pumped into a lower
lying potential surface \cite{Sisyphus89}. In this case kinetic
energy is converted into potential energy which is subsequently
carried away by a spontaneously emitted photon, thereby reducing
the total atomic energy.

\begin{figure}[tb]
\infig{20em}{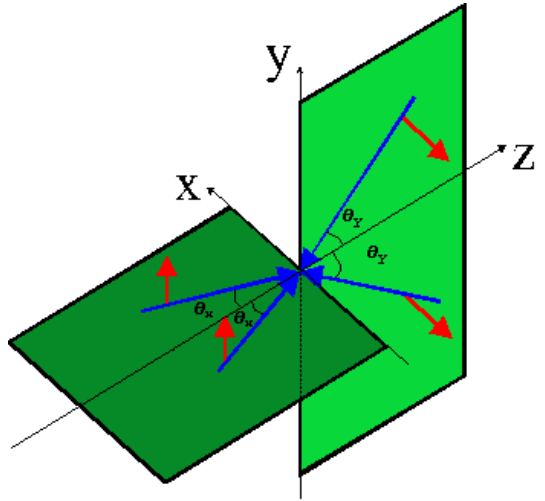}
\caption{Laser beam configuration for the
three-dimensional lin$\bot$lin Sisyphus cooling. Two pairs of
laser beams with crossed linear polarizations induce polarization
and light shift gradients in the three directions of space.}
\label{tetrahedron}
\end{figure}

In this paper we study the so-called 3D-lin$\bot$lin configuration
\cite{3dlinlin}. It is obtained from the standard 1D-lin$\bot$lin
configuration \cite{Sisyphus89} by symmetrically splitting each of
the two laser beams into two parts at an angle $\theta_x$ and
$\theta_y$, respectively, with the ($Oz$) axis in the ($Oxz$) and
($Oyz$) planes respectively. The resulting configuration consists
of two pairs of laser beams in the ($Oxz$) plane and in the
($Oyz$) plane, respectively, as depicted in
Fig.~\ref{tetrahedron}, with orthogonal linear polarizations. An
important property of this configuration in contrast to 3D setups
built of more than four laser beams is that the interference
pattern and thus the topography of the lattice does not change
because of fluctuations of the relative phase between the various
laser beams. Instead, such fluctuations only induce displacements
of the lattice.

%\subsection{The optical potential}

As in most theoretical work, we will consider atoms with a ground
state of angular momentum 1/2 and an excited state of angular
momentum 3/2. Experiments usually fall into the low saturation
regime defined by \be s_0 =
\frac{\Omega_0^2/2}{\Delta^2+\Gamma^2/4}\ll 1, \ee where $s_0$ is
the saturation parameter for an atomic transition with a
Clebsch-Gordan coefficient of one, $\Omega_0$ is the Rabi
frequency for one laser beam, $\Delta$ the detuning of the laser
beams from the atomic resonance frequency, and $\Gamma$ the
natural width of the atomic excited state. This domain is known to
lead to the lowest temperatures. In this situation we may
adiabatically eliminate the Zeeman sublevels of the excited state,
leading to a theory which only involves the ground state sublevels
$|\pm\rangle$ of angular momentum $\pm 1/2$ \cite{Cohen90}.

\begin{figure}[tb]
\infig{20em}{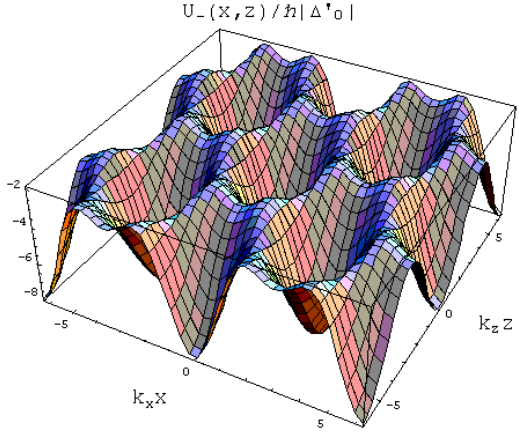} \caption{Section at y=0 of the optical
potential associated with the Zeeman sublevel $|-\rangle$ of the
atomic internal ground state in the case of a negative detuning
($\Delta<0$). $U_{+}$ can be obtained from $U_{-}$ by a
translation of half a spatial period along one of the axes ($Ox$)
and ($Oz$).} \label{optical_potential}
\end{figure}

An atom in state $|\pm\rangle$ then experiences an optical
potential $U_\pm$ given by \bea U_{\pm}(x,y,z) & = &
\frac{8\hbar\Delta_{0}'}{3}\Big[ \cos(k_{x}x)^{2}+\cos(k_{y}y)^{2} \nonumber \\
 & & \mp \cos(k_{x}x) \cos(k_{y}y) \cos(k_{z}z)\Big]
\label{potential} \eea where \be \Delta_{0}'={\Delta}\frac{s_0}{2}
\ee is the light shift per beam and
\begin{mathletters}
\bea
 k_x & = & k\sin({\theta}_x), \\
 k_y & = & k\sin({\theta}_y), \\
 k_z & = & k[\cos\left({\theta}_x)+\cos({\theta}_y\right)],
\eea \label{spatper}
\end{mathletters}
with $k$ the laser wavenumber. The optical potentials are then
periodic in the three directions of space with periods
$\lambda_i=2\pi/k_i$. Equation (\ref{potential}) shows that
$U_\pm(x,y,z)$ has the same functional dependence on $k_x x$ and
on $k_y y$ but a different one on $k_z z$. We will therefore in
the remainder of the paper concentrate on a two-dimensional
subsystem only depending on $x$ and $z$ while fixing $y=0$.
Previous comparisons between 1D and 2D models have shown that the
general behavior of the dynamic variables is the same in different
dimensions but that scaling factors appear \cite{oscillth95}. We
thus expect that the results of our work give physical
interpretations to full 3D laser cooling schemes but exact
numerical values will be changed. Moreover, we will assume a
single lattice angle $\theta=\theta_x=\theta_y$ since this gives
rise to a vanishing mean radiation pressure force in all
directions. The general shape of the optical potential is plotted
in Fig.~\ref{optical_potential}.

In the 3D-lin$\bot$lin configuration, the bottom of each potential
well is harmonic in first approximation with main axis $x$, $y$
and $z$ and with the following frequencies:
\begin{mathletters}
\bea
& & \Omega_{x,y} = 4\sqrt{|\Delta_0'|\omega_r}\frac{k_{x,y}}{k}, \\
& & \Omega_z =
\frac{4}{\sqrt{3}}\sqrt{|\Delta_0'|\omega_r}\frac{k_z}{k}, \eea
where $\omega_R=\hbar k^2/(2M)$ denotes the recoil frequency.
\label{harmonic}
\end{mathletters}
The optical pumping time is \be \tau_p = \frac{9}{8 \Gamma_0'},
\label{optpumping} \ee where $\Gamma_0'=\Gamma s_0 /2$ is the
optical pumping rate. The jumping regime corresponds to a domain
where $\Omega_i{\ll}1/\tau_p$, that is, when an atom undergoes
many optical pumping cycles during a single oscillation or during
a flight over a single potential well. On the contrary, the
oscillating regime corresponds to $\Omega_i{\gg}1/\tau_p$. In this
case, an atom can oscillate or travel over many wells without
undergoing any pumping cycle. Note that in a 3D-lin$\bot$lin
optical lattice, the regimes can be different in different
directions because of the geometrical dependence of the border
between the jumping and the oscillating regimes, \be
\sqrt{\frac{\omega_r}{|\Delta_0'|}}\frac{|\Delta|}{\Gamma}
   \sim \frac{\lambda_i}{\lambda}.
\label{border} \ee

\begin{figure}[tb]
\infig{18em}{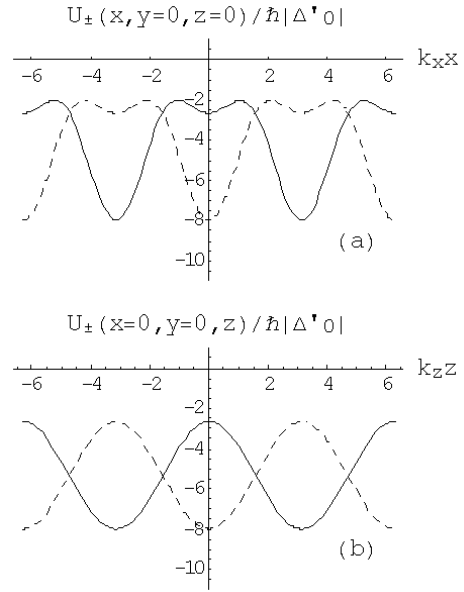} \caption{Sections at (a) $y=z=0$, (b)
$x=y=0$ of the optical potential surfaces $U_{+}$ (solid curve)
and $U_{-}$ (dashed curve).} \label{potential1D}
\end{figure}

The asymmetry between the $x$ and $z$ directions can be seen most
easily in a plot of the optical potentials along the $x$ and $z$
axis, respectively, as shown in Fig.~\ref{potential1D}. These have
different shapes and particularly the crossing between both
potential curves is higher in the transverse direction ($x$) than
it is in the longitudinal one ($z$). As we will see later, this
induces significant differences in the cooling and diffusion
properties.

%\subsection{Semiclassical approach}

As the starting point for the theory presented here we use the
standard Fokker-Planck equation (FPE) of the semiclassical laser
cooling theory \cite{localization96,Castin94,speckle98} where the
external degrees of freedom of the atoms are treated as classical
variables. This is obtained from the Wigner transform
\cite{Wigner32} of the full quantum master equation for external
as well as internal degrees of freedom under the assumption of a
momentum distribution which is much broader than a single photon
momentum, $\Delta P\gg \hbar k$. The FPE for the populations
$\Pi_\pm({\bf r},{\bf p})$ of $|\pm\rangle$ reads \bea &
&\left[\partial_t+\frac{p_i}{M}
\partial_i -(\partial_i U_\pm)
\partial_{p_i}
   \right] \Pi_\pm = \nonumber \\
& &\quad\quad \gamma_\mp\Pi_\mp - \gamma_\pm \Pi_\pm  \nonumber \\
& &\quad\quad -F^i_{\pm\pm} \partial_{p_i} \Pi_\pm
              -F^i_{\mp\pm} \partial_{p_i} \Pi_\mp \nonumber \\
& &\quad\quad +D^{ij}_{\pm\pm} \partial_{p_i}
\partial_{p_j}\Pi_\pm
              +D^{ij}_{\mp\pm} \partial_{p_i} \partial_{p_j}\Pi_\mp.
\label{Fokker} \eea Here $i,j=x,z$ and summation over $i$ and $j$
is assumed. In this equation, ${\gamma}_{\pm}$ is the jumping rate
from the Zeeman sublevel $|\pm\rangle$ to the sublevel
$|\mp\rangle$, $F^i_{\pm\pm}$ represents the radiation pressure
force and $D^{ij}_{\pm\pm}$ the momentum diffusion matrix for
atoms in the internal state $|\pm\rangle$. $F^i_{\pm\mp}$ and
$D^{ij}_{\pm\mp}$ are the corresponding quantities associated with
jumps between different internal states \cite{Petsas99}. Note that
all these coefficients only depend on the atomic spatial position
\cite{Castin94,speckle98}.

A numerical solution of the FPE can be obtained by averaging over
many realizations of the corresponding Lan-gevin equations
\cite{Risken89}. Such a realization consists in following the
trajectory of a single atom which jumps between the two optical
potential surfaces corresponding to the two internal states with
the appropriate probabilities. Between subsequent jumps the atom
experiences potential and radiation pressure forces as well as
random momentum kicks which mimic the momentum diffusion according
to the coefficients of the FPE. We have performed a large number
of such semiclassical Monte-Carlo simulations in order to
investigate the dependence of the steady-state temperature, the
friction coefficient, and in particular the spatial diffusion
coefficient on the various lattice parameters such as detuning
$\Delta$, light shift $\Delta_0'$, and lattice angle $\theta$. We
will discuss these numerical results later in Secs.\
\ref{temperature}, \ref{diffusion}, and \ref{friction}.

%%%%%%%%%%%%%%%%%%%%%%%%%%%%%%%%%%%%%%%%%%%%%%%%%%%%%%%%%%%%%%%%%

\section{Physical picture of spatial diffusion}
\label{physics}

%\subsection{Spatial diffusion process}

\begin{figure}[tb]
\infig{20em}{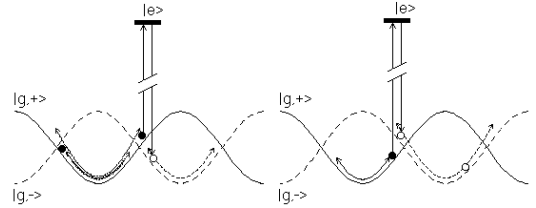} \caption{Processes of spatial
diffusion in an optical lattice. Left: process due to the heating.
Right: process due to optical pumping.} \label{diff_pross}
\end{figure}

As a result of the Sisyphus effect, the atoms are cooled and
trapped in the potential wells and optical lattices are usually
described as atoms well confined in regularly arranged sites (note
that in the lin$\bot$lin lattice, the spatial periods of these
sites are $\lambda_i/2$, the trapping sites corresponding
alternatively to $U_+$ and $U_-$ potential wells). However, in
bright optical lattices the atom confinement is not perfect
because of the strong interaction with the laser light. Two
different processes then produce atomic displacements between
different trapping sites, inducing spatial diffusion (see
Fig.~\ref{diff_pross}). For the sake of simplicity, we describe
these processes in one dimension but they occur analogously in
higher dimensional setups.

On the one hand (see Fig.~\ref{diff_pross} left), a trapped atom
still undergoes fluorescence cycles and thus takes random recoils
due to photon absorption and re-emission. Hence, the oscillating
motion of the atom gets perturbed. Particularly, the atom can
explore regions where its potential energy ($U_\pm$ if the atomic
internal state is $|\pm\rangle$) is not minimum ($U_\pm > U_\mp$).
In such regions, optical pumping cycles preferentially transfer
the atom into the lower potential curve and the atom is cooled and
trapped in the neighboring potential well (elementary Sisyphus
cooling process). This process induces atomic transfers from a
site to a neighboring one in another potential curve.

On the other hand (see Fig~\ref{diff_pross} right), when a trapped
atom oscillates in a potential well, it has a small but non zero
probability of being optically pumped into the upper potential
curve. In the jumping regime the atom is immediately pumped back
into its initial trapping potential well. This effect thus induces
heating and noisy oscillating trajectory of the atom leading
indirectly to spatial diffusion via transfers between neighbouring
potential wells. On the contrary, in the oscillating regime the
atom is not immediately re-pumped and travels over several
potential wells before undergoing an elementary Sisyphus cooling
process again which traps it into another potential well.

The diffusion process linked to optical pumping is much more
efficient than the one due to recoils except for very small laser
detunings \cite{Cohen90}. We will thus focus on the second process
to describe spatial diffusion in periodic multi-dimensional
optical lattices. We will see that the differences of this process
in the jumping and the oscillating regimes induce a dramatic
difference in the behavior of the spatial diffusion coefficients.

%\subsection{Random walk models}

In a simple model, we can describe the diffusive behavior of the
atomic cloud as random walks of atoms between periodic trapping
sites \cite{Itzykson91}. Let us assume that an atom is trapped  in
one specific potential well and jumps after a time $\tau$ to
another well. The spatial diffusion coefficient in the direction
$i \in \{x,y,z\}$ is \be D_{Si}=\frac{d_{i}^2}{2\tau}
\label{walk_lattice} \ee where $d_i$ is the mean free path in
direction $i$.

In the jumping regime, as discussed above, an atom essentially
transfers from a trapping site to a neighboring one and thus
$d_i\sim\lambda_i$. The life time of an oscillatory external state
is on average of the order of $1/\Gamma_0'$ \cite{JYC-GG92},
independently of $\lambda_i$. Hence, $\tau \sim 1/\Gamma_0'$ and
\be D_{Si} \sim \lambda_i^2 \Gamma_0'. \label{diff_jumping} \ee

In the oscillating regime, an atom travels over several lattice
sites before it is trapped again. Here $d \sim \overline{v} \tau$
where $\overline{v} \simeq \sqrt{k_B T/M}$ is the average velocity
with $M$ the atomic mass and $\tau$ the time of one flight. As it
will be justified in Secs.~\ref{theory} and \ref{temperature},
$k_B T$ is proportional to $\hbar|\Delta_0'|$ independently of
$\Delta$ and $\theta$. $\tau$ is of the order of $1/\Gamma_0'$
again\cite{JYC-GG92}. Hence, \be D_S \sim \frac{\hbar}{M}
\frac{|\Delta|}{\Gamma}. \label{diff_oscillating} \ee Note that we
implicitly assume straight line flights and we do not consider
anisotropic effects. Obviously, because of the potential shape and
anisotropy, a dependence of $D_{Si}$ on $\lambda_x$, $\lambda_y$
and $\lambda_z$ should be added in Eq.~(\ref{diff_oscillating})
but our simple model does not provide its determination.
Nevertheless, in the case of a strong anisotropy ($\lambda_i \ll
\lambda_j$), the setup is almost one-dimensional in the
i-direction so it is expected that $D_{Si}$ does not depend on
$\lambda_i$. Indeed, at large space scale, the length of one
flight of a particle moving on a 1D-periodic potential is
independent of the periodicity.

We want to emphasize that these discussions are only valid for
lattice parameters far away from the domain of {\it d\'ecrochage}
for spatial diffusion. For small potential depths, rare long
flights dominate the diffusion which therefore becomes anomalous
\cite{anomalous96}.

%%%%%%%%%%%%%%%%%%%%%%%%%%%%%%%%%%%%%%%%%%%%%%%%%%%%%%%%%%%%%%%%%

\section{Brownian motion model}
\label{theory}

Before turning to a detailed discussion of the numerical results,
we will now derive a simple analytical model which will help to
understand the main features of the cooling scheme.

The basic idea of this model is to consider the atomic dynamics as
a Brownian motion, as it has been successfully applied to Doppler
cooling \cite{Cohen90} and 1D Sisyphus cooling in the jumping
regime \cite{Sisyphus89,Cohen90}, but taking some properties of
localization into account \cite{Munich,Gatzke97}. The fundamental
ingredients are therefore the derivation of an average friction
force and an average momentum diffusion coefficient.

Let us consider an atom at position ${\bf r} = x{\bf e}_x + z{\bf
e}_z$ moving with a constant velocity ${\bf v} = v_x{\bf e}_x +
v_z{\bf e}_z$. In this case the FPE (\ref{Fokker}) reduces to \bea
(\partial_t + {\bf v}\nabla) \Pi_\pm
   &=&  -\gamma_\pm\Pi_\pm + \gamma_\mp\Pi_\mp \nonumber \\
   &=& -(\gamma_-+\gamma_+)\Pi_\pm + \gamma_\mp
\label{eq:th1} \eea where we used $\Pi_++\Pi_-=1$. The jump rates
are \be \gamma_{\pm} = \frac{4}{9}\Gamma_0'(1+\cos^2k_xx \pm 2
\cos k_xx\cos k_z z). \ee Expanding the populations in powers of
the velocity in the form $\Pi_\pm = \sum_n \Pi_\pm^{(n)}$ and
inserting into Eq.~(\ref{eq:th1}), yields the stationary solutions
\begin{mathletters}\bea
\Pi_\pm^{(0)} & = & \frac{\gamma_\mp}{\gamma_-+\gamma_+} \\
\Pi_\pm^{(n)} & = & Q^n \Pi_\pm^{(0)} \eea\end{mathletters} with
the operator \be Q = -\frac{1}{\gamma_-+\gamma_+} {\bf v}\nabla.
\label{eq:q} \ee Formally, the velocity and position dependent
level populations can thus be written as \be \Pi_\pm =
\frac{1}{1-Q}\Pi_\pm^{(0)}. \ee The total force averaged over the
internal atomic states is then given by \be {\bf F}({\bf r},{\bf
v}) = -(\Pi_+ \nabla U_+ + \Pi_- \nabla U_-). \ee In order to
derive a space and velocity averaged friction force in the form
${\bf F}=-\left(v_x\alpha_x{\bf e}_x + v_z\alpha_z{\bf
e}_z\right)$ we now have to make certain assumptions on the
stationary atomic distribution.

In 1D laser cooling one usually assumes a flat spatial
distribution of the atoms in the lattice. However, as we have seen
in Sec.~\ref{sisyphus}, the shape of the 3D optical potential
differs significantly in the different directions. In particular
the potential barrier between neighboring potential wells is much
higher in the transverse than in the longitudinal direction. From
Eq.~(\ref{potential}) and Fig.~\ref{potential1D} we see that in
the $x$ direction the potential depth is of the order of
$16\hbar|\Delta_0'|/3$ whereas we will see later that the
steady-state temperature is of the order of $2\hbar|\Delta_0'|$.
Therefore we expect strong localization in that direction. On the
contrary, in the $z$ direction the potential depth is of the order
of $8\hbar|\Delta_0'|/3$ and the atoms will be less localized.
Instead of a flat spatial distribution of the atoms we will thus
assume a thermal distribution \be P({\bf r}) \propto
   \exp\left\{-\frac{\gamma_-U_+ +\gamma_+U_-}{k_BT(\gamma_+ + \gamma_-)}
       \right\}
\label{spatdistr} \ee corresponding to the optical potential
averaged over the internal atomic state and for a given, yet
unknown, temperature $T$. Let us emphasize that assuming a thermal
distribution is not a priori justified for laser cooled samples
but show significant quantitative deviations from such a simple
behavior \cite{Gerz93,Molmer94}. In fact, our numerical
simulations give actual spatial distributions in qualitative
agreement with Eq.~\ref{spatdistr} but show significant
quantitative deviations from such a simple behavior. However, we
only use this approximation here to obtain a qualitative
understanding of the exact results obtained numerically and we
will see later that our results derived here are in good
quantitative agreement with the simulations.

Because of the symmetry of $P({\bf r})$, only terms containing odd
powers of the velocity in $\Pi_\pm$ contribute to the averaged
force. We may thus restrict ourselves to \be \Pi_\pm^{odd} =
\sum_n \Pi_\pm^{(2n+1)} = \frac{1}{1-Q^2} \Pi_\pm^{(1)}. \ee As a
further simplification we will now also average over velocity and
therefore replace $Q^2$ by \be \langle Q^2\rangle =
     -\frac{2\omega_R k_B T/\hbar}{\langle(\gamma_-+\gamma_+)^2\rangle}
      [(k_x/k)^2 + (k_z/k)^2]
\label{eq:thq2} \ee where $\langle...\rangle$ is the spatial
average with respect to $P({\bf r})$. Equation (\ref{eq:thq2}) is
obtained by substituting $v_i\rightarrow 0$, $v_i^2\rightarrow k_B
T/M$, $\partial_i^2\rightarrow -k_i^2$ ($i=x,z$) in
Eq.~(\ref{eq:q}). From this we finally get the friction
coefficients \bea \alpha_i = -\frac{6\frac{\Delta}{\Gamma} \hbar
k_i^2 C_i}{1-\langle Q^2\rangle} \label{eq:thal} \eea with
\begin{mathletters}\bea
C_x & = & \left\langle\frac{\sin^4(k_x x)\cos^2(k_z
z)}{[1+\cos^2(k_x x)]^3}
          \right\rangle, \\
C_z & = & \left\langle\frac{\cos^2(k_x x)\sin^2(k_z
z)}{[1+\cos^2(k_x x)]^2}
          \right\rangle.
\eea\end{mathletters}

Along the lines of \cite{JDCCTold} and using the same
approximations as for the friction, we derive averaged momentum
diffusion coefficients \be D_{pi} =
\frac{8\frac{\Delta}{\Gamma}\Delta_0' \hbar^2 k_i^2 C'_i}
              {1-\langle Q^2\rangle}
\label{eq:thdp} \ee with
\begin{mathletters}\bea
C'_x & = & \left\langle 4\Pi_+^{(0)}\Pi_-^{(0)}
      \frac{\sin^2(k_x x)\cos^2(k_z z)}{1+\cos^2(k_x x)}
   \right\rangle, \\
C'_z & = &\left\langle 4\Pi_+^{(0)}\Pi_-^{(0)}
      \frac{\cos^2(k_x x)\sin^2(k_z z)}{1+\cos^2(k_x x)}
   \right\rangle.
\eea\end{mathletters} For simplicity we have not written a term of
the momentum diffusion which arises from the random recoil of
absorbed and spontaneously emitted photons. This varies as
$\Gamma_0'$ and thus can be neglected for large detuning
($|\Delta|>\Gamma$).

In a model of Brownian motion, the steady-state temperature then
fulfills \be k_B T = \frac{1}{2}\left(\frac{D_{px}}{\alpha_x} +
        \frac{D_{pz}}{\alpha_z}\right),
\label{Taver} \ee that is, averaging Eq.~(\ref{TDpalpha}) over the
$x$ and $z$ directions. The right hand side is obtained from
Eqs.~(\ref{eq:thal}) and (\ref{eq:thdp}) which themselves depend
on the temperature via Eq.~(\ref{spatdistr}). Thus,
Eq.~(\ref{Taver}) yields an implicit equation for $k_B
T/\hbar|\Delta'_0|$ which can be solved numerically, e.g., by
iteration of Eqs.~(\ref{spatdistr})-(\ref{Taver}) recursively.
Note that no lattice parameter, such as the lattice angle or the
laser detuning, appears in this equation. $T$ is thus strictly
proportional to $\hbar|\Delta_0'|$ and independent of $\theta$ and
$\Delta$. We find \be k_B T = 1.545 \,\hbar |\Delta_0'|.
\label{tempnum} \ee This temperature can then be used to determine
the following numerical values: \be
\begin{array}{l l}
C_x = 0.0356, \quad & C_z = 0.0747, \\
C'_x = 0.0409,\quad & C'_z = 0.0874,
\end{array}
\label{coeffC}\ee and \be \langle(\gamma_-+\gamma_+)^2\rangle =
2.55 \Gamma_0^{\prime 2}. \ee

Finally, we want to derive an approximate expression for the
spatial diffusion coefficients. To this end, we must again take
the atomic localization into account. While Eq.~(\ref{TDpalpha})
for the relation between temperature, momentum diffusion
coefficient and friction coefficient approximately holds for
trapped and untrapped atoms, the corresponding equation
(\ref{DsTalpha}) for the spatial diffusion only holds for free
atoms. Indeed a cloud of completely trapped atoms achieves a
stationary spatial distribution and hence shows no spatial
diffusion. Using Eq.~(\ref{tempnum}) and the assumption of a
thermal momentum and spatial distribution, we calculate
numerically that a fraction of 55.6\% of all atoms have a total
energy above the potential depth along $z$, and a fraction of
15.3\% above the potential depth along $x$ as discussed before.
Taking only these free atoms into account, we finally obtain
effective spatial diffusion coefficients, which correspond to
those observed in the numerical simulations or in actual
experiments, in the form \bea
D_{Sx} & = & 0.554 \frac{\hbar}{M}\left(\frac{k}{k_x}\right)^2 \nonumber \\
   & & \times
   \left\{\frac{|\Delta_0'|}{\omega_R}\frac{\Gamma}{|\Delta|} +
      1.21\frac{|\Delta|}{\Gamma}[(k_x/k)^2 + (k_z/k)^2]
   \right\}, \nonumber \\
D_{Sz} & = & 0.956 \frac{\hbar}{M}\left(\frac{k}{k_z}\right)^2 \nonumber \\
   & & \times
   \left\{\frac{|\Delta_0'|}{\omega_R}\frac{\Gamma}{|\Delta|} +
      1.21\frac{|\Delta|}{\Gamma}[(k_z/k)^2 + (k_x/k)^2]
   \right\}.
\label{diffnum} \eea This expression is in good qualitative
agreement with our physical discussion (see Sec.~\ref{physics}) in
both the jumping and the oscillating regime. We will further
discuss this point in Sec.~\ref{diffusion}.

%%%%%%%%%%%%%%%%%%%%%%%%%%%%%%%%%%%%%%%%%%%%%%%%%%%%%%%%%%%%%%%%%

\section{Steady-state kinetic temperature}
\label{temperature}

We performed a systematic study of the temperature and the spatial
diffusion as a function of the lattice parameters, exploring a
large domain containing both the jumping and the oscillating
regime. More precisely, we performed numerical semi-classical
Monte-Carlo simulations with the following parameters: \bea & &
\theta = 15^\circ,\, 30^\circ,\, 45^\circ,\, 60^\circ,\,
75^\circ,\,
   \nonumber \\
& & \Delta/\Gamma = -2,\, -3,\, -5,\, -10,\, -15,\, -20,\, -25,\,
-30,\,
   \nonumber \\
& & \Delta_0'/\omega_r = -150,\, -300,\, -450,\, -600,\, -750.
\nonumber \eea

In this and the following sections we present the results of the
simulations and compare them with the analytical model discussed
above. All of our discussions and conclusions rely on the complete
set of data, even if the figures only contain a few sample curves
for the sake of clarity.

We first study the atomic cloud steady-state temperature resulting
from the competition between slowing and heating processes. We
calculate the average square velocity over the whole atomic cloud
at each time step of the simulations. The temperature in direction
$i$ is then given by \be k_B T_i = M \langle v_i^2 \rangle.
\label{numtemp} \ee A second method to obtain the temperature
numerically is to fit a Gaussian to the simulated momentum
distribution. We have checked that the widths of these Gaussians
indeed give the same temperatures as those obtained from
Eq.~(\ref{numtemp}). For a broad initial velocity distribution,
the temperature first decreases in time (thermalization phase),
but finally reaches a steady state.

The thermalization time lies between $50/\Gamma_0'$ and
$100/\Gamma_0'$ for large enough lattice angles ($\theta \geq
30^\circ$ in the simulations), but for $\theta = 15^\circ$ it is
approximately $400/\Gamma_0'$ in the transverse direction. In this
case, the spatial period is large and an atom needs to fly a long
time to undergo efficient Sisyphus cooling. Such an increase of
the cooling time is also observed in the longitudinal direction
for large angles but is less important because $\lambda_z$ does
not reach very large values.

\begin{figure}[tb]
\infig{20em}{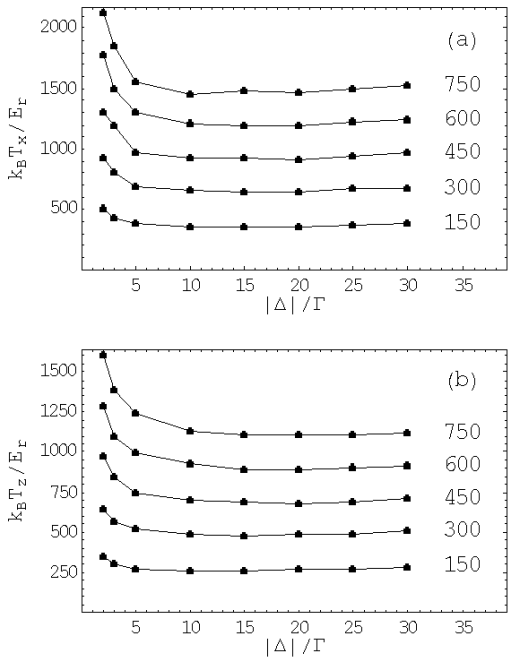} \caption{Transverse (a) and longitudinal (b)
steady-state temperatures versus the laser detuning $\Delta$ for a
fixed lattice angle $\theta = 45^\circ$ and for various light
shifts (the values of $|\Delta_0'|/\omega_r$ are quoted to the
right of the corresponding curves).} \label{TvD}
\end{figure}

Both the transverse and the longitudinal temperatures are shown in
Fig.~\ref{TvD} versus the laser detuning for a fixed lattice angle
and for several light shifts per beam. This exhibits two domains
where the temperature behaves differently. For small detunings
($|\Delta|<10\Gamma$) we find a rapid decrease of the temperature
with increasing $|\Delta|$, whereas for large detunings the
temperature is independent of $\Delta$ but increases approximately
linearly with $\Delta_0'$. This agrees well with the general form
\be \frac{k_B T_i}{\hbar |\Delta_0'|} = \left[A_i + B_i
\left(\frac{\lambda_i}{\lambda}\right)^2\right]
\left(\frac{\Gamma}{|\Delta|}\right)^2 +C_i, \label{Texact} \ee
where $A_i$, $B_i$ and $C_i$ ($i = x,z$) are numerical factors,
as, for example, has been found in Ref.~\cite{speckle98}. The
first term of Eq.~(\ref{Texact}) has not been found in
Sec.~\ref{theory} because in the momentum diffusion,
Eq.~(\ref{eq:thdp}), we neglected the term due to absorption and
spontaneous emission.

%\subsection{Regime of temperature independent of the laser detuning}

\begin{figure}[tb]
\infig{20em}{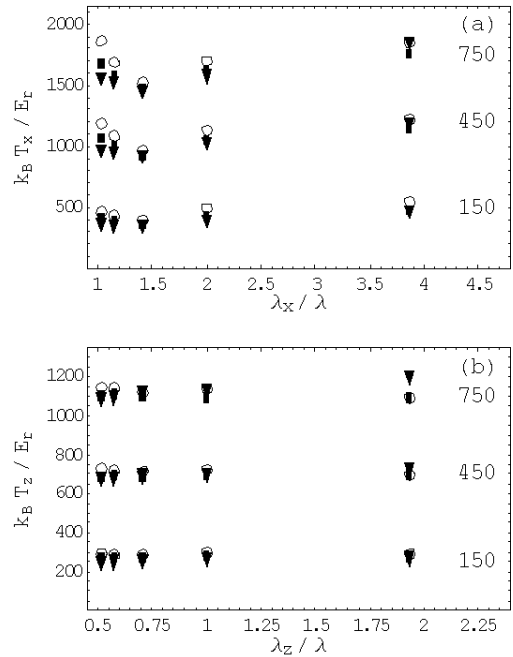} \caption{Transverse (a) and longitudinal
(b) temperatures versus the corresponding lattice period for
various laser detunings (triangles: $\Delta=-10\Gamma$, squares:
$\Delta=-20\Gamma$, circles: $\Delta=-30\Gamma$) and light shifts
(the values of $|\Delta_0'|/\omega_r$ are quoted to the right of
the corresponding curves).} \label{Tvtheta}
\end{figure}

Let us first concentrate on the oscillating regime,
$|\Delta|\gg\Gamma$. In Fig.~\ref{Tvtheta} we plot the temperature
versus the lattice spatial period for various laser detunings and
light shifts in this domain. We find that the temperature is
nearly independent of the lattice angle, i.e., of the spatial
periods. The temperature is thus strictly linear in the potential
depth and independent of any other lattice parameter as predicted
by Eq.~(\ref{tempnum}). Such a property has been observed
experimentally \cite{Carminati01} but we emphasize that our result
here holds for a broader range of parameters in the oscillating
regime as well as in the jumping regime. A linear fit to the
numerical results gives
\begin{mathletters}
\bea
k_B T_x & \simeq & 2 \hbar|\Delta_0'|+123.3 \hbar\omega_r, \\
k_B T_z & \simeq & 1.43 \hbar|\Delta_0'|+62.6 \hbar\omega_r. \eea
\label{kBT}
\end{mathletters}

In the range of parameters investigated here, these values of the
temperatures agree with Eq.~(\ref{tempnum}) with an accuracy of
about $10\%$.

Equations~(\ref{kBT}) show that the temperature is anisotropic,
$T_x>T_z$, which is in good agreement with experimental
observations by A.\ Kastberg and coworkers \cite{Kastberg01}. This
is a consequence of the asymmetry between the transverse and
longitudinal directions for the optical potential in the
lin$\bot$lin lattice. In the physical picture of Sisyphus cooling
\cite{Sisyphus89,Cohen90},  cooling ends once an atom is trapped
in a single potential well and hence the steady-state temperature
is proportional to the potential depth. As already discussed in
Sec.~\ref{theory} this is about twice as large in the transverse
direction as in the longitudinal.  Hence, the transverse
temperature would be expected to be about twice as large as the
longitudinal. However, correlations between these two directions
tend to equilibrate the temperatures and in the simulations we
therefore find $T_x$ to be only about $1.4$ times larger than
$T_z$.

%\subsection{Divergence in temperature}

We now turn to the jumping regime, $|\Delta|<10\Gamma$. We find a
completely different behavior of the temperature. This is a
consequence of the increasing contribution of absorption and
spontaneous emission to the force fluctuations \cite{Carminati01}.
Physically, an atom experiences many photon recoils during one
elementary cooling process and is thus more likely to escape from
the trapping potential. Hence, the atom can reach a steady state
temperature larger than the potential well depth.

\begin{figure}[tb]
\infig{20em}{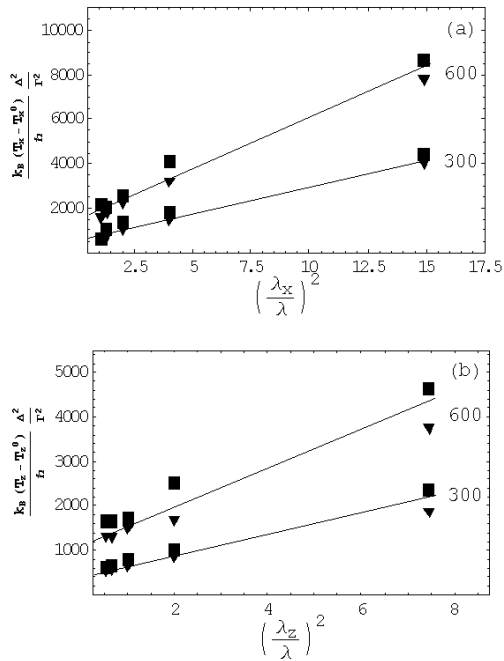} \caption{Variation of the (a)
transverse and (b) longitudinal temperature as a function of the
lattice spatial periods for small detunings. $T_i^0$ denotes the
temperature in the domain of large $\Delta$, $\Delta/\Gamma=-2$
(triangles), $\Delta/\Gamma=-3$ (squares), and
$|\Delta_0'|/\omega_r =300$, $600$. The lines are linear fits of
the form of Eq.~\ref{Texact}.} \label{Tdecrotheta}
\end{figure}

In Fig.~\ref{Tdecrotheta} we plot the increase of the temperature
compared to the oscillating regime as a function of the lattice
periods $\lambda_i$. In excellent agreement with
Eq.~(\ref{Texact}) we find that $T_i$ is proportional to
$\Delta'_0$ and to $\lambda_i^2$.

%%%%%%%%%%%%%%%%%%%%%%%%%%%%%%%%%%%%%%%%%%%%%%%%%%%%%%%%%%%%%%%%%

\section{Spatial diffusion of the atomic cloud}
\label{diffusion}

We now turn to the study of the spatial diffusion of the atomic
cloud. In the simulations we calculate the average square position
$\langle r_i^2\left(t\right)\rangle$ in each direction over the
whole cloud.

In the thermalization phase the hot atoms follow almost ballistic
trajectories and the cloud expands rapidly. For longer times the
expansion reaches a normal diffusion regime where \be \langle
r_i^2\left(t\right)\rangle = 2 D_{S i} t + \Delta r_{i,0}^2
\label{normal} \ee and $\Delta r_{i,0}^2$ is a constant depending
on the initial space and velocity distribution. Note that $D_{S
i}$ does not depend on this initial distribution as we verified in
the simulations.

\begin{figure}[tb]
\infig{20em}{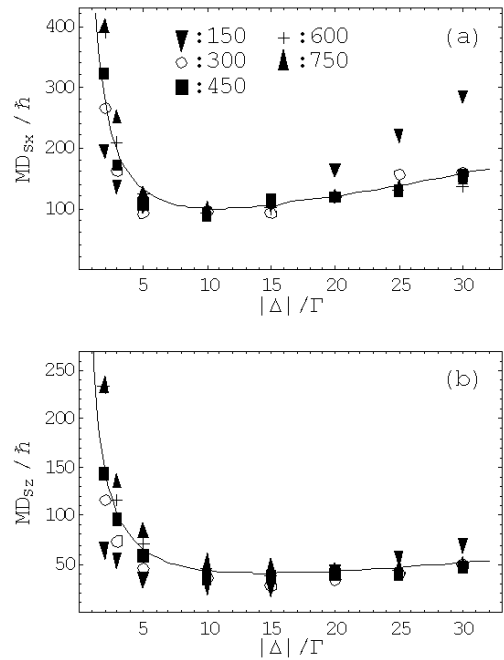} \caption{Transverse (a) and longitudinal (b)
spatial diffusion coefficients as a function of $\Delta$ for
various $\Delta_0'$ (the values of $|\Delta_0'|/\omega_r$ are
plotted in the graphs) and for $\theta = 45^\circ$. The line
corresponds to the analytical fit, Eq.~(\ref{coeffdiff}), for
$\Delta_0'=-450\omega_r$.} \label{DvD}
\end{figure}

In Fig.~\ref{DvD} we plot the transverse and longitudinal spatial
diffusion coefficients versus the lattice detuning for various
light shifts and for a given lattice angle. Figure~\ref{DvD}
clearly shows two domains where the spatial diffusion coefficient
behaves differently. For small detunings, $D_{Si}$ decreases
rapidly with $|\Delta|$ and increases with $|\Delta_0'|$, whereas
for large detunings, $D_{Si}$ increases with $|\Delta|$ and does
not depend on $\Delta_0'$ except for $\Delta_0'=-150 \omega_r$.
The latter corresponds to a relatively shallow potential and the
system is close to the transition to anomalous diffusion
\cite{oscillth95,anomalous96,Carminati01}.

Fitting the numerical results with an expression of the form of
Eq.~(\ref{diffnum}), we find: \bea
& & D_{Sx} = 0.50 \frac{\hbar}{M}\left(\frac{k}{k_x}\right)^2 \nonumber \\
   & & \times
   \left\{\frac{|\Delta_0'|}{\omega_R}\frac{\Gamma}{|\Delta|} +
      1.73\frac{|\Delta|}{\Gamma}[1.16(k_x/k)^2 + 0.83(k_z/k)^2]
   \right\}, \nonumber \\
& & D_{Sz} = 1.30 \frac{\hbar}{M}\left(\frac{k}{k_z}\right)^2 \nonumber \\
   & & \times
   \left\{\frac{|\Delta_0'|}{\omega_R}\frac{\Gamma}{|\Delta|} +
      1.00\frac{|\Delta|}{\Gamma}[0.83(k_z/k)^2 + 1.16(k_x/k)^2]
   \right\}. \nonumber \\
& & \label{coeffdiff} \eea The coefficients of this fit are in
good quantitative agreement with the analytical result
(\ref{diffnum}). The main difference is the factor of 1.39 which
amounts to the difference between the longitudinal and the
transverse temperatures as found in Sec.~\ref{temperature}.
Figure~\ref{DvD} shows that Eq.~(\ref{coeffdiff}) yields an
excellent fit to the numerical data. From Eq.~(\ref{border}) and
Eqs.~(\ref{diffnum}), (\ref{coeffdiff}) we observe that the two
domains where $D_S$ behaves differently correspond to the jumping
and the oscillating regimes.

%\subsection{Jumping regime}

\begin{figure}[tb]
\infig{20em}{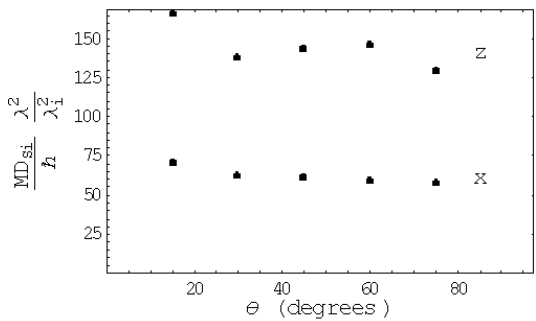} \caption{Transverse (x) and
longitudinal (z) spatial diffusion coefficients divided by the
corresponding square spatial period $\lambda_i^2$ as a function of
$\theta $ for $\Delta_0'=-750 \omega_r$ and $\Delta=-3\Gamma$
(jumping regime).} \label{Dvthetajump}
\end{figure}

In the jumping regime we find:
\begin{mathletters}
\bea
D_{Sx} & = & 0.025\, \lambda_x^2 \Gamma_0', \\
D_{Sz} & = & 0.066\, \lambda_z^2 \Gamma_0',
\eea\label{coeffdiffjump}\end{mathletters} in good qualitative
agreement with our physical discussion in Sec.~\ref{physics}, see
Eq.~(\ref{diff_jumping}). In particular, $D_{Si}$ is proportional
to $\lambda_i ^2$ as shown in Fig.~\ref{Dvthetajump} and
proportional to the optical pumping rate $\Gamma_0'$. The
different values of the numerical coefficients indicate that the
trapping is stronger in the transverse direction than in the
longitudinal, in agreement with our discussion in
Sec.~\ref{theory}.

%\subsection{Oscillating regime}

\begin{figure}[tb]
\infig{20em}{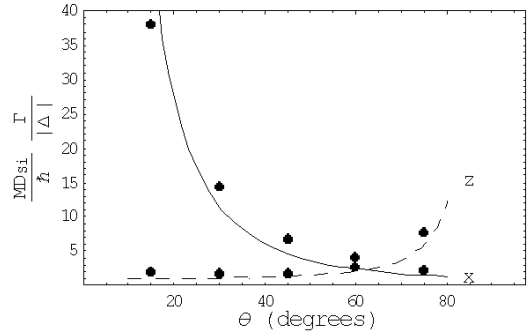} \caption{Transverse (x) and
longitudinal (z) spatial diffusion coefficients divided by
$|\Delta|$ as a function of $\theta $ for $\Delta=-30\Gamma$ and
$\Delta_0'=-300 \omega_r$. The points correspond to the numerical
simulations and the lines correspond to Eq.~(\ref{coeffdiffosc})
(oscillating regime).} \label{Dvthetaosc}
\end{figure}

On the contrary, in the oscillating regime we find:
\begin{mathletters}
\bea D_{Sx} & = & \frac{\hbar}{M}
   \left[1 + 0.72 \left(\frac{\lambda_x}{\lambda_z}\right)^2\right]
   \frac{|\Delta|}{\Gamma} \\
D_{Sz} & = & 1.08\frac{\hbar}{M}
   \left[1+1.39\left(\frac{\lambda_z}{\lambda_x}\right)^2\right]
   \frac{|\Delta|}{\Gamma}.
\eea \label{coeffdiffosc}
\end{mathletters}
In this regime, $D_S$ is proportional to $\Delta$ as expected from
Eq.~(\ref{diff_oscillating}) and the angular dependence is
qualitatively given by Eq.~(\ref{diffnum}). In
Fig.~\ref{Dvthetaosc} we show this angular dependence in the
oscillating regime. Here, the transverse and longitudinal
directions are not independent because the potential wells
significantly deflect the trajectories of atoms travelling over
many optical potential wells.  The dependence of $D_{Si}$ on
$\theta$ thus contains both $\lambda_x$ and $\lambda_z$. This is
dramatically different to the situation in the jumping regime
where atoms only jump between adjacent wells. Equation
(\ref{coeffdiffosc}) also confirms that for $\lambda_i
\ll\lambda_{j}$, $D_{Si}$ does not depend on $\lambda_i$ as
predicted in Sec.~\ref{physics}.

%%%%%%%%%%%%%%%%%%%%%%%%%%%%%%%%%%%%%%%%%%%%%%%%%%%%%%%%%%%%%%%%%

\section{Friction force}
\label{friction}

The theoretical model of Sec.~\ref{theory} was based on the
description of the atomic dynamics by a Brownian motion model. Let
us now further investigate the validity of such a description by
testing in the numerical simulations some characteristics of
Brownian motion. We particularly perform a direct numerical
measurement of the friction coefficients and test the validity of
the Einstein relation (\ref{DsTalpha}).

%\subsection{A Brownian motion}

In order to probe the atomic dynamics we submit the atoms to a
constant, space and velocity independent force in addition to the
forces due to the atom-light interaction. In an experiment this
could be provided simply by gravity or, for example, by the
radiation pressure force of an additional weak laser beam. In a
Brownian motion model such a constant force ${\bf F}^{(c)} =
F^{(c)}_x {\bf e_x}+F^{(c)}_z{\bf e_z}$ will give rise to a
constant mean velocity  $\langle{\bf v}\rangle = v_x {\bf
e_x}+v_z{\bf e_z}$ of the atomic cloud with \be \langle v_i
\rangle = \frac{F^{(c)}_i}{\alpha_i}. \label{velocity} \ee Because
of the linearity of the Brownian equations of motion the kinetic
temperature and the spatial diffusion coefficients are not
changed.

Adding such a constant force in the numerical simulations along
the $i$-direction, we observed in fact that the atomic cloud
experiences a drift in this direction at a constant velocity. In
the ideal case of a pure Brownian motion any amplitude of ${\bf
F}^{(c)}$ can be used, but in the case of Sisyphus cooling the
friction coefficient is velocity dependent. In order to get $v_i$
proportional to $F^{(c)}_i$, it is thus essential to use a small
enough force which induces a global drift much smaller than the
width of the velocity distribution. Under this condition the
temperature and the spatial diffusion do not depend on ${\bf
F}^{(c)}$ and Eq.~(\ref{velocity}) can be used to numerically find
unique friction coefficients $\alpha_i$.

\begin{figure}[tb]
\infig{20em}{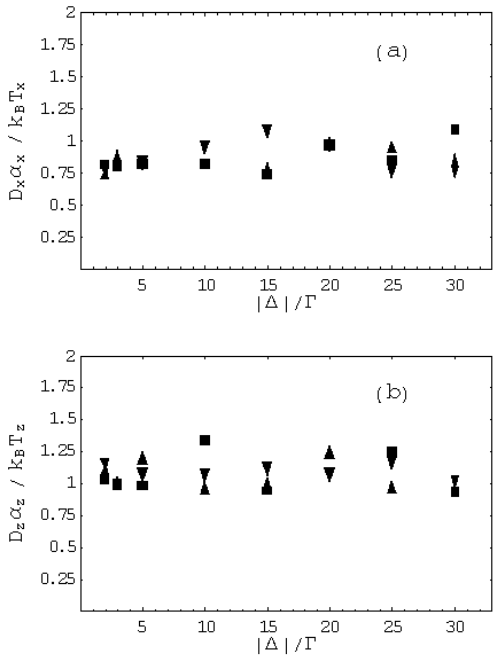} \caption{Ratio of the friction
coefficients calculated with Eq.~(\ref{velocity}) and with
Eq.~(\ref{DsTalpha}) for various $\Delta$ and for
$|\Delta_0'|/\omega_r = 150$, $300$, $450$.} \label{relation}
\end{figure}

We can then compare these numerical results with the friction
coefficient obtained via the Einstein relation (\ref{DsTalpha})
using the numerically found values of the temperature (see
Sec.~\ref{temperature}) and of the spatial diffusion coefficient
(Sec.~\ref{diffusion}). In Fig.~\ref{relation} we plot the ratio
of these two friction coefficients and find deviations of about
15\%. Hence, the dynamics of an atomic cloud in an optical lattice
is in reasonable agreement with a two-dimensional Brownian motion
model. Note that the eigen-directions of the motion are the $x$
and $z$ directions in good agreement with the theoretical model.

In order to better understand this result let us briefly return to
the model developed in Sec.~\ref{theory}. As has been shown in the
previous section, the numerically obtained diffusion coefficient
agrees well with the analytical result  (\ref{diffnum}). In the
derivation of the latter we have assumed that the cloud of atoms
can be split into a trapped fraction and into a free fraction.
Only the free atoms were taken into account for the spatial
diffusion. Similar arguments must also be considered for the
friction coefficient. Adding a {\it small} constant force will
leave a trapped particle in its initial potential well, and
therefore the fraction of trapped atoms does not contribute to the
mean velocity $\langle {\bf v}\rangle$ of the cloud. Thus, the
measured friction coefficient using Eq.~(\ref{velocity}) should be
given by the analytical result (\ref{eq:thal}) divided by the
fraction of free atoms. Hence, for the measured values of the
spatial diffusion and of the friction both sides of
Eq.~(\ref{DsTalpha}) are corrected by the same factor. In other
words, the Einstein relation holds because the measured quantities
only involve the freely travelling atoms for which a Brownian
motion model works well.

%\subsection{The friction coefficient versus the lattice parameters}

\begin{figure}[tb]
\infig{20em}{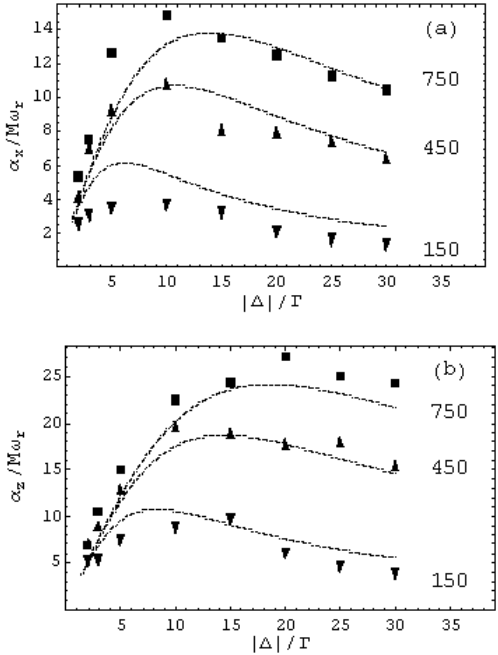} \caption{Transverse (a) and longitudinal (b)
friction coefficients versus $\Delta$ (data points) and the curves
corresponding to Eq.~(\ref{frictioncoeff}) for $\theta=45^\circ$
and for $\Delta_0'/\omega_r = -150, -450, -750$.} \label{GvD}
\end{figure}

Therefore, we obtain an analytic fit to the measured friction
coefficient by inserting Eqs.~(\ref{kBT}) and (\ref{coeffdiff})
into Eq.~(\ref{DsTalpha}). We find
\begin{mathletters}
\bea \alpha_x & \simeq & \frac{2 \hbar
k_x^2\frac{|\Delta|}{\Gamma}}
   {1+\left(k_x^2+0.72k_z^2\right)\frac{\Delta^2}{\Gamma^2}
      \frac{\hbar}{M|\Delta_0'|}} \\
\alpha_z & \simeq & \frac{0.55 \hbar k_z^2\frac{|\Delta|}{\Gamma}}
   {1+0.4\left(k_z^2+1.39k_x^2\right)\frac{\Delta^2}{\Gamma^2}
      \frac{\hbar}{M|\Delta_0'|}}.
\eea \label{frictioncoeff}
\end{mathletters}
These approximate expressions are compared with the numerically
obtained values of the friction coefficient in Fig.~\ref{GvD}. We
see that there is excellent agreement both qualitatively and
quantitatively.

The behavior of $\alpha_i$ is again different in the jumping and
the oscillating regime. In the jumping regime, $\alpha_i$ is
proportional to $|\Delta|/\Gamma$ and approximately independent of
$\Delta_0'$ \cite{Sisyphus89,Cohen90}:
\begin{mathletters} \bea \alpha_x & = & 2 \hbar k_x^2
\frac{|\Delta|}{\Gamma}
\\ \alpha_z & = & 0.55 \hbar k_z^2
\frac{|\Delta|}{\Gamma}. \eea
\label{frictioncoeffjump}\end{mathletters} However, Fig.~\ref{GvD}
exhibit a small dependence of $\alpha_i$ versus $\Delta_0'$ in the
jumping regime and this is not forecasted by the model. In fact,
the kinetic temperature is not proportional but linear in
$\Delta_0'$ and this induce a dependence of the spatial
distribution $P$ in $\Delta_0'$. Coefficients $C_i$ are thus
$\Delta_0'$-dependent and this can explain the discrepancy.

In the oscillating regime, $\alpha_i$ is proportional to
$\Gamma_0'=\Gamma\Delta_0'/\Delta$ and depends on both $\lambda_x$
and $\lambda_z$ as discussed in Sec.~\ref{diffusion}:
\begin{mathletters} \bea \alpha_x & = & \frac{2
M}{1+0.72k_z^2/k_x^2} \frac{\Gamma \Delta_0'}{\Delta}
\\ \alpha_z & = & \frac{1.38 M}{1+1.39k_x^2/k_z^2} \frac{\Gamma
\Delta_0'}{\Delta}. \eea \label{frictioncoeff}\end{mathletters}
The expression found in the simulations is in good qualitative
agreement with the expression derived in the theoretical model
Eq.~(\ref{frictioncoeff}). Note however that coefficients $C_i$
are not given by Eq.~\ref{coeffC} but are to be calculated
considering the free atoms which contribute to spatial diffusion
only.

%%%%%%%%%%%%%%%%%%%%%%%%%%%%%%%%%%%%%%%%%%%%%%%%%%%%%%%%%%%%%%%%%

\section{Conclusions}
\label{conclusions}

In conclusion, we performed a systematic study of the behavior of
an atomic cloud in a lin$\bot$lin optical lattice with the help of
semi-classical Monte-Carlo simulations. We explored a broad range
of lattice parameters including the jumping and the oscillating
regime.

The temperature was found to be linear in the potential depth and
independent of the laser detuning and of the lattice angle in a
broad range of parameters. We have shown that the temperature is
anisotropic, the transverse one being larger than the longitudinal
one by a factor of $1.4$. All these results are well explained
with the help of the physical picture of Sisyphus cooling and are
in good agreement with experimental results.

The spatial difusion $D_S$ was studied in the regime of normal
diffusion. The behavior of $D_S$ differs significally in the
jumping and in the oscillation regimes. In the first, $D_S$
decreases with $|\Delta|$ and increases with $|\Delta_0'|$. In the
second, $D_S$ increases linearly with $|\Delta|$ and does not
depend on $|\Delta_0'|$. The behavior of $D_S$ as a function of
the lattice spatial periods is also different in both regimes:
whereas $D_{Si}$ is proportional to $\lambda_i^2$ in the jumping
regime, it is a function of $\lambda_x$ as well as $\lambda_z$ in
the oscillating regime. This reveals correlations between the
transverse and longitudinal directions of the lattice.

By adding a constant force in the Monte-Carlo simulations we could
numerically measure the friction coefficients and we showed that
the Einstein relations are fulfilled. This supports a description
of the dynamics in terms of Brownian motion. The friction
coefficient  $\alpha_i$ is proportional to $\Delta$ and
$1/\lambda_i^2$ in the jumping regime. In the oscillating regime,
$\alpha_i$ is proportional to $\Delta_0'$ and $1/\Delta$, and the
dependence on the lattice geometry involves both $1/\lambda_x$ and
$1/\lambda_z$.

The numerical results have been found to be in good agreement with
a simple theoretical model based on a semi-classical approach. We
derive the steady-state temperature, the friction force, and the
spatial diffusion from a model of Brownian motion taking into
account atomic localization in the optical potential wells. To
explain the measured friction and spatial diffusion, the atomic
cloud must be split into a trapped part and a free part. While in
general both parts contribute to the internal and external
dynamics, only the free fraction of atoms is responsible for the
observed expansion of the cloud and for the drift of the center of
mass under the influence of a weak constant force.

The spatial diffusion of atomic clouds in optical lattices is
usually studied in pump-probe spectroscopy experiments, using the
properties of the Rayleigh line \cite{JYC-GG96}. However, the
validity of this method has never been proven. We expect that the
models discussed here will serve to this verification by providing
a systematic theoretical study of the directly measured spatial
diffusion coefficients.

Finally, it should be noted that our restriction to $\theta_x =
\theta_y$ has been guided by experimental restrictions but is not
necessary. Indeed, when this condition is not fulfilled, the
unbalanced radiation pressure induces a fast escape of the atomic
cloud from the optical lattice which makes experimental
investigations difficult. Nevertheless, this situation could be of
great interest because it offers the opportunity of studying
optical lattices with different spatial periods along the $x$ and
$y$ axes and interesting anisotropic effects could be found.

%%%%%%%%%%%%%%%%%%%%%%%%%%%%%%%%%%%%%%%%%%%%%%%%%%%%%%%%%%%%%%%%%

\acknowledgments

We are indebted to Yvan Castin for numerous enlightening
discussions. We also thank Anders Kastberg's group for the
communication of their experimental results before publication.
Laboratoire Kastler-Brossel is an unit\'e de recherche de l'Ecole
Normale Sup\'erieure et de l'Universit\'e Pierre et Marie Curie
associ\'ee au Centre National de la Recherche Scientifique (CNRS).
This work was partially supported by the European Commission (TMR
network ``Quantum Structures'', contract FMRX-CT96-0077) and the
Austrian Science Foundation FWF (project P13435-TPH and SFB
``Control and Measurement of Coherent Quantum Systems'').

\end{document}